\documentclass[pre]{revtex4}
\usepackage{amssymb} 
	\usepackage{amsfonts}
	\usepackage{epsfig}

\begin{document}
\title{Behavior of collective variables in complex nonlinear stochastic models of finite size}
\author{M. Morillo and J. M. Casado}
\address{Area de F\'{\i}sica Te\'orica. Universidad de Sevilla.\\
Apartado Correos 1085, 41080 Sevilla (Spain)}

\date{\today}

\begin{abstract}
We consider the behavior of a collective variable in a complex system formed by a finite number of interacting subunits. Each of them is characterized by a degree of freedom with an intrinsic nonlinear bistable stochastic dynamics. The lack of ergodicity of the collective variable requires the consideration of a feedback mechanism of the collective behavior on the individual dynamics. We explore numerically this issue within the context of two simple finite models with a feedback mechanism of the Weiss mean-field type: a global coupling model and another one with nearest neighbors coupling.
\end{abstract}

\maketitle
\section{Introduction}
Complex hierarchical systems admit several levels of
description. At a microscopic level, the interest lies on the
dynamical behavior of the constituent elements. Thus, at this level of
description, one needs to follow the time evolution of very many
microscopic variables. On the other hand, at the macroscopic level,
one describes the system in terms of just a few collective variables
characterizing it as a whole. The laws describing the behavior of
collective variables could, in principle, be deduced
from the dynamical laws governing the elementary constituents.
In practice this is not an easy task as, in general, the dynamical equations for the collective variables 
can not be obtained from the microscopic dynamics without further approximations.  
On the other hand, 
in general, when studying a complex systems, there are just the collective
variables which are accessible to experimental measurements. The detailed dynamics
of the constituent elements has to be inferred from the information about the 
collective variables. 

A fundamental property common to all complex systems made of a
collection of interacting elements (subsystems) is the existence of
feedback mechanisms of the whole system behavior into the dynamics of the 
subsystems \cite{KOMETANI1975}.
The evolution of the system is determined by
those of its elements in a statistical way and, at the same time, the
feedback loops put the subsystems under the control of the global
system. Such feedback loops, for example, give rise to the biological
laws that characterize the phenomenon of life. In fact, this mutual
interlevel control introduces a self-regulation mechanism to the
dynamics of the hierarchical system against external forces and
boundary conditions.

In this work, we have studied two very simple models describing a finite system 
composed of several subsystems. Each of them is characterized by a single stochastic variable 
$X_i(t),\; i=1,\cdots,N$. The subsystems have an intrinsic nonlinear bistable dynamics and 
they are coupled among them either in a global or local form. The whole system is characterized 
by a collective variable. We will assume that the information about the system comes 
from the observation of the collective variable. The randomness of the subsystem variables
is due to a noise term in their dynamics. We will take this noise to be a white Gaussian noise
characterized by a noise parameter $D$. This noise can originate for instance, from the weak interactions between the subsystems and
a large thermal bath with a temperature $D/k_B$.

A goal of this work is to study the form of the distribution
function $P_{eq}(s)= \lim_{t\to\infty} \langle \delta(s-S(t) \rangle $
for the collective variable and to compare it with the distribution
function of a single element $P_{eq}(x)$ for small finite systems. 
As we will see, in the absence of feedback, both distributions are unique, regardless of the initial preparation of the system. This is a consequence of the fact that the joint probability distribution satisfies a linear Fokker-Planck equation. If the observations of the collective variable are compatible with such a behavior, then we can accept the microscopic description. 

But, if one observes that, depending upon the noise value and the strength of the subsystems coupling, the equilibrium distribution might not be unique and it might depend on the initial preparation of the system, then the linear Fokker-Planck equation (or equivalently, the set of coupled Langevin-like equations) is not adequate. To deal with such a situation, we propose stochastic dynamics leading to nonlinear Fokker-Plank equations. The idea is simply an extension of the old Weiss mean-field approach to the equilibrium properties of magnetic systems. Then, as the system parameters are varied, the dynamics leads us from just a single stationary solution to situations where two distributions are stable and the system goes to one or the other depending upon the initial conditions.  

We should point out that such a non-uniqueness of a single variable distribution function was studied long ago by several authors \cite{DESAI1978,DAWSON1983,SHIINO}. They were able to derive a nonlinear Fokker-Planck equation for a single variable in the limit of an infinite system. We emphasize that, in this work, we are interested in small, finite systems. 

The intrinsic bistable nonlinearity of the subsystems dynamics forbids explicit analytical solutions, but it is essential for the behavior indicated above. A closed, explicit, stochastic evolution equation for the collective variable can not be found from the subsystems dynamics.  We will then rely on numerical simulations of the stochastic evolution equations for the subsystems variables. 

\section{Two stochastic nonlinear models}
\subsection{A model with global interactions}
We consider a system formed by a set of $N$ subsystems each one described by a single
stochastic variable $X_{i}$, whose dynamics is given by the
coupled Langevin equations (in dimensionless form)
\begin{equation}
\dot{X}_{i}(t)=X_{i}(t)-X_{i}^{3}(t)+\frac{\theta}{N}\sum_{j=1}^{N}\Big (X_{j}(t)-X_{i}(t)\Big )+\xi_{i}(t),\qquad
(i=1,2,\ldots, N),
\label{EQ001}
\end{equation}
where $\theta$ is the strength of the coupling and $\xi_{i}(t)$ are Gaussian white noises with
\begin{equation}
\langle\xi_{i}(t)\rangle=0;\qquad \langle\xi_{i}(t)\xi_{j}(s)\rangle=2D\delta_{ij}\delta(t-s).
\label{EQ002}
\end{equation}   
This model was introduced some years ago by Kometani and Shimizu
\cite{KOMETANI1975} in a biophysical context and was later analyzed by
Desai and Zwanzig \cite{DESAI1978} and Dawson \cite{DAWSON1983} from a
more general statistical mechanical perspective. An alternative
formulation can be casted in terms of the linear Fokker-Planck
equation for the joint probability distribution
$f_{N}(x_{1},x_{2},\ldots,x_{N},t)$,
\begin{equation}
\frac{\partial P_{N}}{\partial t}=\sum_{i=1}^{N}\frac{\partial
}{\partial x_{i}}\Big(\frac{\partial U}{\partial
  x_{i}}P_{N}\Big)+ D \sum_{i=1}^{N}\frac{\partial^{2}P_{N}}{\partial
  x^{2}_{i}},
\label{FPE}
\end{equation}
where $U$ is the potential energy relief,
\begin{equation}
U=\sum_{i=1}^{N}V(x_{i})+\frac{\theta}{4N}\sum_{i=1}^{N}\sum_{j=1}^{N}(x_{i}-x_{j})^{2},
\end{equation}
with the single particle potential
\begin{equation}
V(x)=-\frac{x^{2}}{2}+\frac{x^{4}}{4}.
\end{equation}

\begin{figure}
\centerline{\epsfig{figure=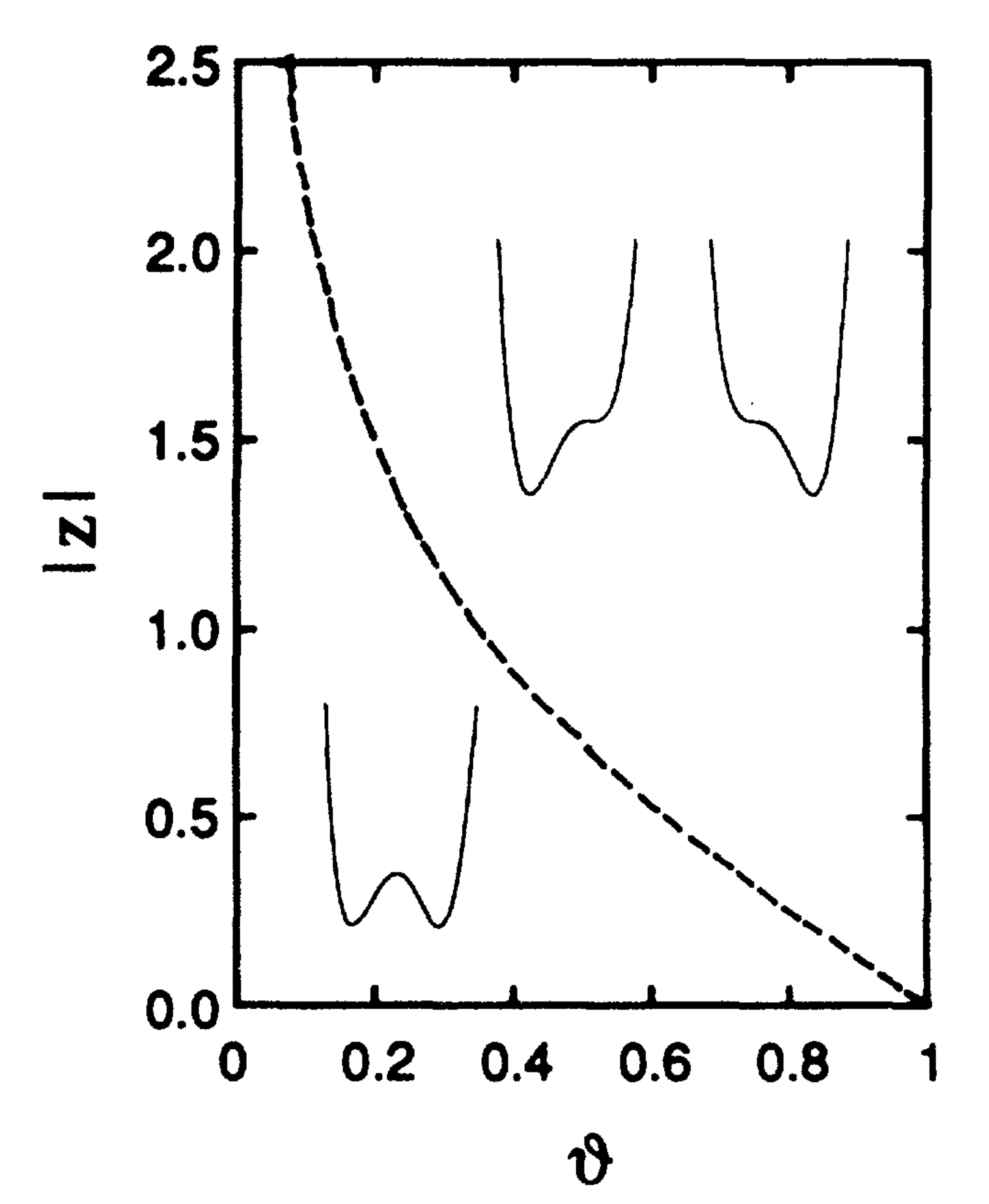,height=6cm}}
\caption{\label{FIG1}\small Bifurcation line for the DZ model with
  $0<\theta<1$. The variable $|z|=|\theta-1|/\sqrt{2D}$ has been used
  instead of $D$. Under the line there is a unique, symmetric,
  one-particle equilibrium distribution function that is
  bistable. Above it there are two equilibrium probability
  distributions with nonzero mean value.}
\end{figure}

In the limit $N\to\infty$, one can write a
Nonlinear Fokker-Planck Equation (NLFPE) for
the one particle probability distribution, $P_{1}(x,t)$ \cite{DESAI1978,DAWSON1983,SHIINO, FRANK,KURSTEN2016}
\begin{equation}
\frac{\partial P_{1}(x,t)}{\partial t}=\frac{\partial}{\partial x}\Big\{\big[(\theta-1)x+x^{3}-\theta\langle x(t)\rangle\big]P_{1}(x,t)\Big\}+D\frac{\partial^{2}P_{1}(x,t)}{\partial x^{2}},
\label{langinf}
\end{equation}
where
\begin{equation}
\label{EQ006}
\langle x(t)\rangle=\int_{-\infty}^{\infty}x P_{1}(x,t)dx.
\end{equation}
 In this limit, the system undergoes an order-disorder phase
 transition signaled by a change in the form of the equilibrium
 one-particle probability distribution $P_{eq}(x)$ and by the fact
 that for some values of the parameters $\theta$ and $D$, the
 ergodicity is broken and there are more than one stable equilibrium
 distribution. The system ends up in one of them depending on its
 initial preparation. In Fig. \ref{FIG1} we sketch different regions
 in the parameter space $0<\theta<1$ and $|z|=|\theta-1|/\sqrt{2D}$,
 separated by a transition line. Below the transition line, there is
 just one single distribution, regardless of the initial preparation
 of the system, leading to a single equilibrium $P_{eq}(x)$ with two
 maxima. This is indicated in the figure by drawing a double well
 potential. On the other hand, above the transition line, two stable
 monomodal equilibrium distributions are possible. Depending upon the
 initial condition the system relaxes to one or the other. This
 feature is illustrated in the figure by drawing two asymmetrical
 single minima potentials.

Let us define a collective variable
\begin{equation}
S(t)=\frac{1}{N}\sum_{j=1}^{N}X_{j}(t),
\label{EQ007}
\end{equation}
Then, the Langevin equations (\ref{EQ001}) can be casted in the equivalent form
\begin{equation}
\dot{X}_{i}(t)=(1-\theta)X_{i}(t)-X_{i}^{3}(t)+\theta S(t)+\xi_{i}(t),\qquad (i=1,2,\ldots, N).
\label{langfin}
\end{equation} 
The nonlinearity of equations (\ref{langfin}) prevent us from writing a
closed Langevin equation for $S(t)$. Note that there is a feedback of the collective
dynamics in the Langevin dynamics of each degree of freedom. But, as we analyzed in a previous work
\cite{GOMEZ2009}, with this type of feedback for the finite model 
the single variable distribution is unique for any set of parameter values. This is consistent with the fact that the joint
probability distribution $P(x_1,\cdots,x_N,t)$ satisfies a linear
Fokker-Plank equation with a unique distribution function for any
values of the parameters.

We now propose to replace the model with  
 \begin{equation}
\dot{X}_{i}(t)=(1-\theta)X_{i}(t)-X_{i}^{3}(t)+\theta \langle S(t)\rangle+\xi_{i}(t),\qquad (i=1,2,\ldots, N),
\label{EQ100}
\end{equation}
Here, $\langle S(t)\rangle$ represents a noise average of the
collective variable defined in Eq.\ (\ref{EQ007}). Clearly, the
Langevin equation, Eq.\ (\ref{EQ100}) has to be solved concurrently
with the evaluation of this average. The proposal amounts to
replace in the dynamical evolution of each individual degree of
freedom the whole stochastic process $S(t)$ by its noise average, in
the spirit of Weiss mean-field approximation to magnetism. Note that 
within this model, 
 the collective variable is itself a fluctuating 
quantity, even though just its average value influences the subsystems dynamics.  

\subsection{A model with nearest neighbors interactions}
This model is described by the (dimensionless) set of Langevin equations
\begin{equation}
\dot{X}_{i}(t)=X_{i}(t)-X_{i}^{3}(t)+\theta \Big ( X_{i+1}+X_{i-1} \Big )+\xi_{i}(t),\qquad (i=1,2,\ldots, N),
\label{NN1}
\end{equation}
with the periodic conditions $ X_0 =X_N$ and $X_{N+1}=X_1$. The noises
$\xi(t)$ are white Gaussian with the properties given in
Eq. (\ref{EQ002}). As in the global interaction model, the joint
probability distribution function satisfies a linear Fokker-Planck
equation similar to the one in Eq. (\ref{FPE}) but with
the potential energy relief,
\begin{equation}
U=\sum_{i=1}^{N}V(x_{i})-\theta \sum_i x_i \Big ( x_{i-1}+x_{i+1} \Big ),
\end{equation}
Then, also for this case,  the joint
probability distribution $P(x_1,\cdots,x_N,t)$ satisfies a linear
Fokker-Plank equation with a unique distribution function for any
values of the parameters. 

As in the global interaction case, we will now replace Eq. (\ref{NN1}) with
\begin{equation}
\dot{X}_{i}(t)= X_i -X_i^3 + 2 \theta \langle S(t)\rangle+\xi_{i}(t).
\label{EQ200}
\end{equation}
As said above, in both situations, even after using the Weiss approach, the collective variable $ S(t) $ for $N$ finite  
is still a random process with a probability distribution defined by
\begin{equation}
P(s,t) = \int dx_1 dx_2 \cdots dx_N\; \delta \Big (s-\frac 1N \sum x_i \Big) P_N(x_1,x_2,\cdots,x_N,t) 
\label{pst}
\end{equation}
 
 Notice that within the mean field dynamics, the joint probability distribution
 $P_N(x_1,x_2,\cdots,x_N,t) $ factorizes as a product of single particle
 distributions, each of them satisfying a NLFPE similar to the one in
 (\ref{langinf}) for the single particle distribution in the infinite
 size limit. \\

Defining the characteristic function $ \Gamma(\mu) $ of a distribution $P(y)$ as
$$
\Gamma(\mu) = \int dy\, e^{i\mu y} P(y)
$$
one can easily see that the generating function
of the collective variable and that of a single particle variable are
related by
$$
\Gamma_s(\mu) = \int ds\, e^{i\mu s} P(s) = \prod_{i=1}^N \Gamma_i(\frac \mu N) 
$$
Thus, the corresponding cumulant generating functions are related by 
 $G_s(\mu) = \sum_{i=1}^N G_i(\frac \mu N) $. Then, the cumulants $C_{s,n}(t) $
associated to the stochastic process $S(t)$, and those associated to the $x(t)$ process, $C_{x,n}(t) $ are
related by  
\begin{equation} 
C_{s,n}(t)=\left .\frac{\partial G_s(\mu)}{\partial \mu} \right |_{\mu=0} = \frac {C_{x,n}(t)}{N^{n-1}}.
\label{cum}
\end{equation}
In the limit of very large $N$,  $P(s)$ becomes a very narrow  
peak around its first moment as indicated in \cite{DESAI1978,DAWSON1983,SHIINO}.

\section{Numerical procedure}
We have solved numerically the corresponding Langevin equations,
Eqs. (\ref{langfin},\ref{NN1}) and 
Eqs.\~ (\ref{EQ100},\ref{EQ200}) for systems with a small number of subsystems using
very many noise realizations. The numerical method used has been
detailed elsewhere \cite{our}. After a convenient relaxation time so
that the equilibrium probability distributions have been reached,
we construct histograms
that approximate the equilibrium probability distributions
$P_{eq}(s)$ and $P_{eq}(x)$.

Notice that the generation of Langevin trajectories for all the
degrees of freedom for the mean field dynamics requires the knowledge of $\langle S(t) \rangle $. Thus, at each time step during the evolution, this
noise average is evaluated from the instantaneous values of each
subsystem variable for each realization of the noise.  Namely,
\begin{equation}
\langle S(t)\rangle=\frac{1}{{\cal N}_{T}}\sum_{\alpha=1}^{{\cal
    N}_{T}}\Big(\frac{1}{N}\sum_{j=1}^{N}X_{i}^{(\alpha)}(t)\Big),
\end{equation}
where ${\cal N}_{T}$ is the number of noise realizations considered.

\begin{figure}
\centerline{\epsfig{figure=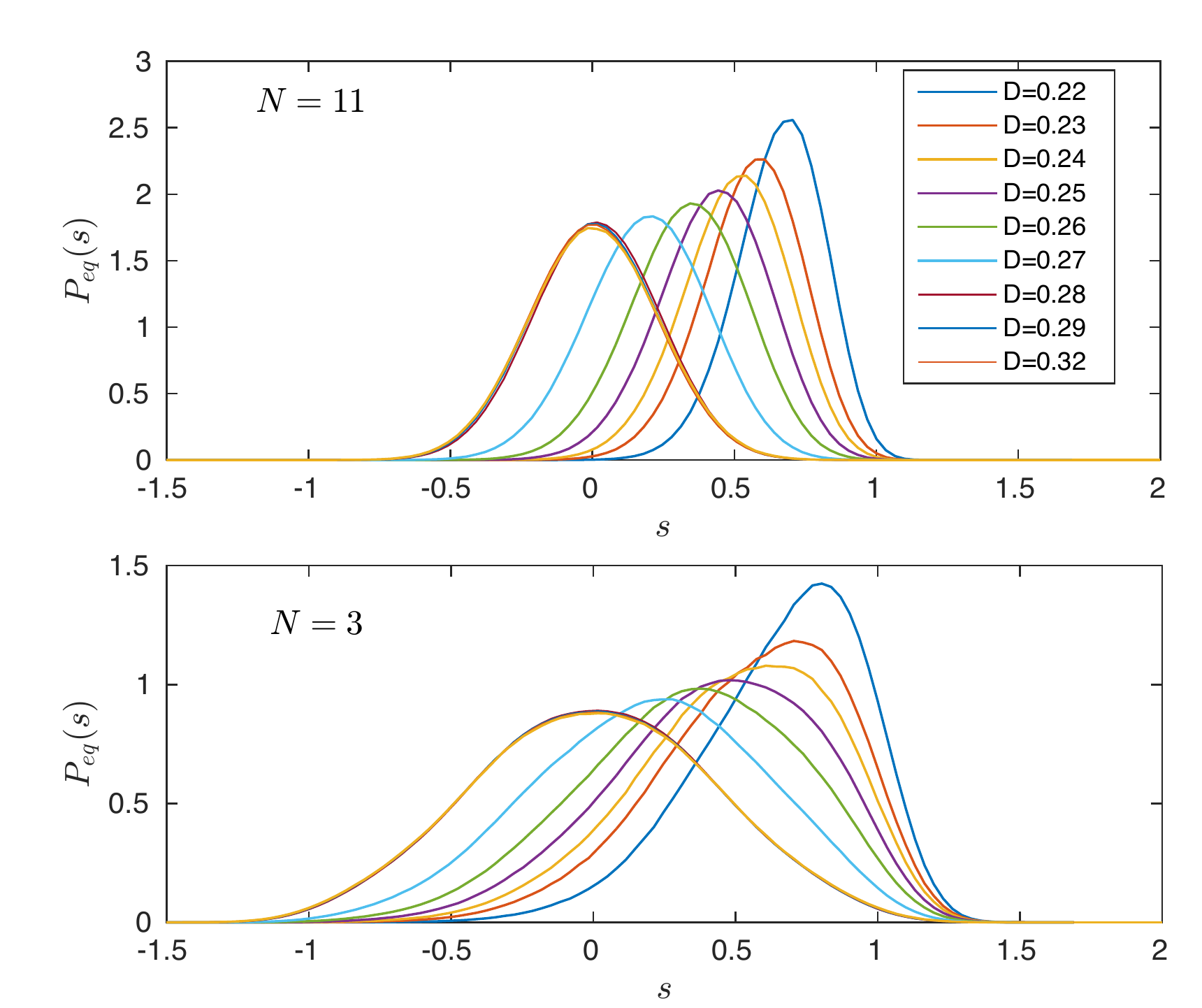,height=6cm}}
\caption{\label{ps}\small Equilibrium distribution $P_{eq}(s)$ for two
  systems with $N=11$ (upper panel) and $N=3$ (lower panel)
  for several values of the moise strength $D$. $\theta=0.5$}
\end{figure}

Let us first consider the case of global coupling with mean-field dynamics. In Fig. \ref{ps} we depict the equilibrium distribution of the
collective variable $P_{eq}(s)$ for several values of the noise
strength and two system sizes,
$N=11$ (upper panel) and $N=3$ (lower panel). In all cases, we have
used $\theta=0.5$ and initial conditions such that $S(0)=0.3$. As seen in the figure, the
qualitative behavior is quite independent of
$N$, except for the narrowing of the distribution as $N$ increases. For the parameter values considered,
$P_{eq}(s)$ is always monomodal, with a peak that is displaced from $s=0$ to
higher values of $s$ as the noise is decreased. Note that the peak is
located at positive values of $s$ because, at the initial time, we
located the value of $S(0)$ to be positive. Had we started from
the same initial values, but with a negative sign, the plots for
$P_{eq}(s)$ would be as in Fig. \ref{ps}, but with the
peaks located at negative $s$ values. 

\begin{figure}
\centerline{\epsfig{figure=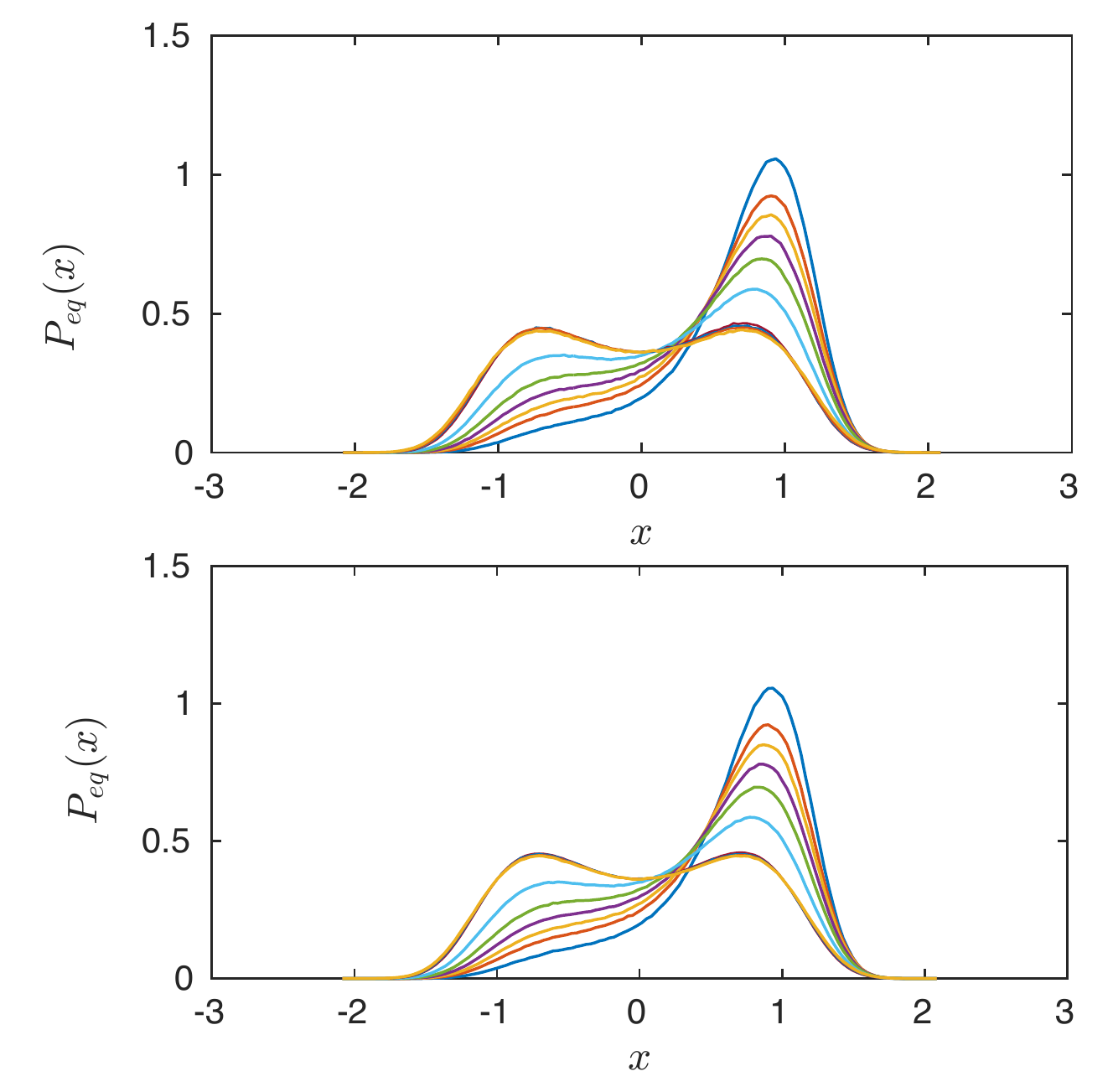,height=6cm}}
\caption{\label{px}\small Equilibrium distribution $P_{eq}(x)$ for two
  systems with $N=11$ (upper panel) and $N=3$ (lower panel)
  for several values of the noise strength $D$. $\theta=0.5$}
\end{figure}

In Fig. \ref{px}, we depict the equilibrium distribution for a single
variable $P_{eq}(x)$ for several values of the noise strength and for
two systems with  $N=11$ (upper panel)
and $N=3$ (lower panel). In all cases, we have used $\theta=0.5$. Again, the
qualitative behavior is quite independent of $N$. By contrast with the behavior of $P_{eq}(s)$, the single variable distribution
$P_{eq}(x)$ changes its shape
from a bimodal distribution centered at $x=0$ to a monomodal one, with
a peak that is displaced to nonzero values of $x$ as the noise is
decreased. 

\begin{figure}
\centerline{\epsfig{figure=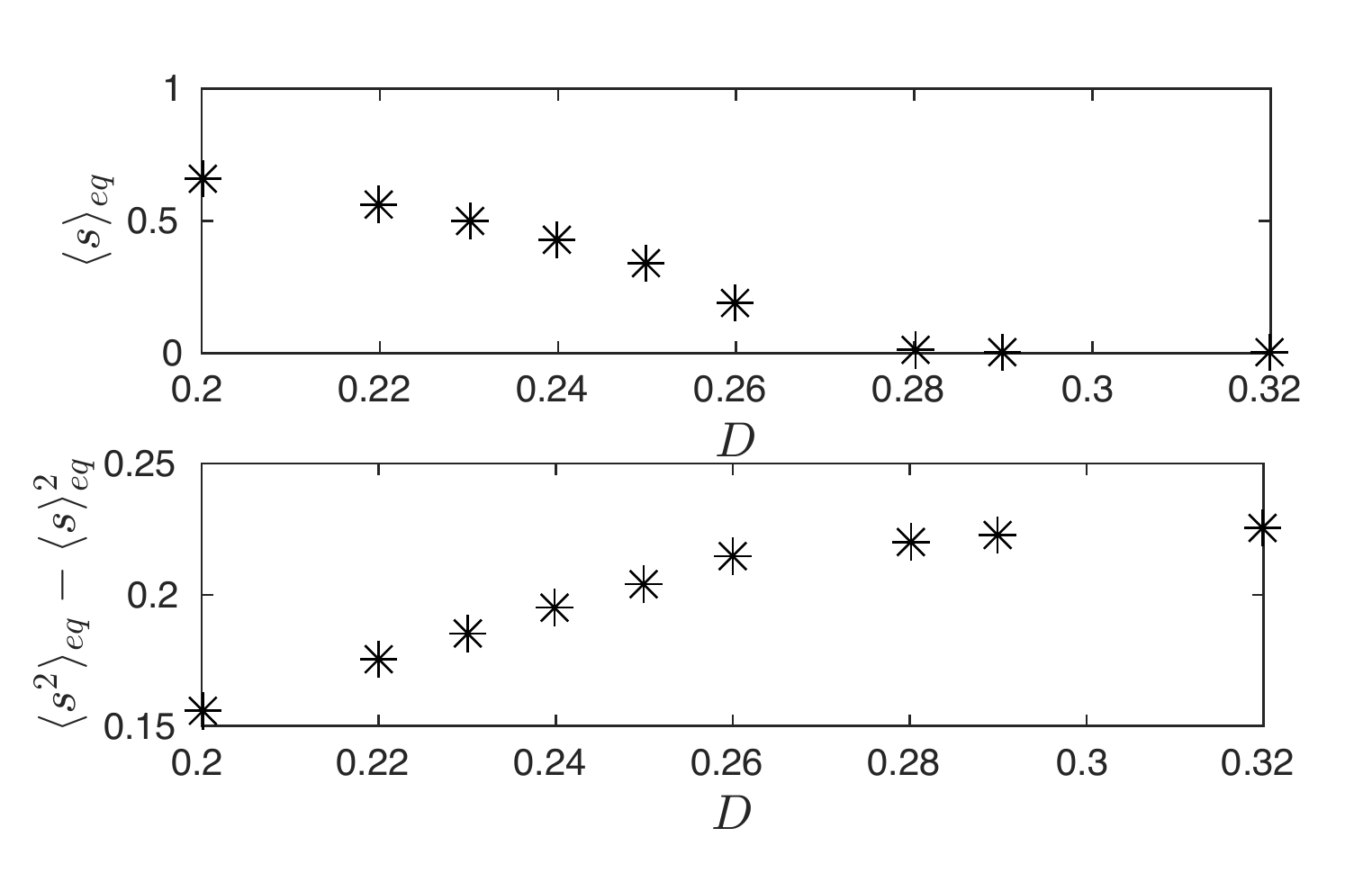,height=6cm}}
\caption{\label{sm}\small The first two equilibrium cumulants of the collective variable for 
$\theta=0.5$ and a range of noise values $D$, for a system with $N=11$. 
}
\end{figure}

\begin{figure}
\centerline{\epsfig{figure=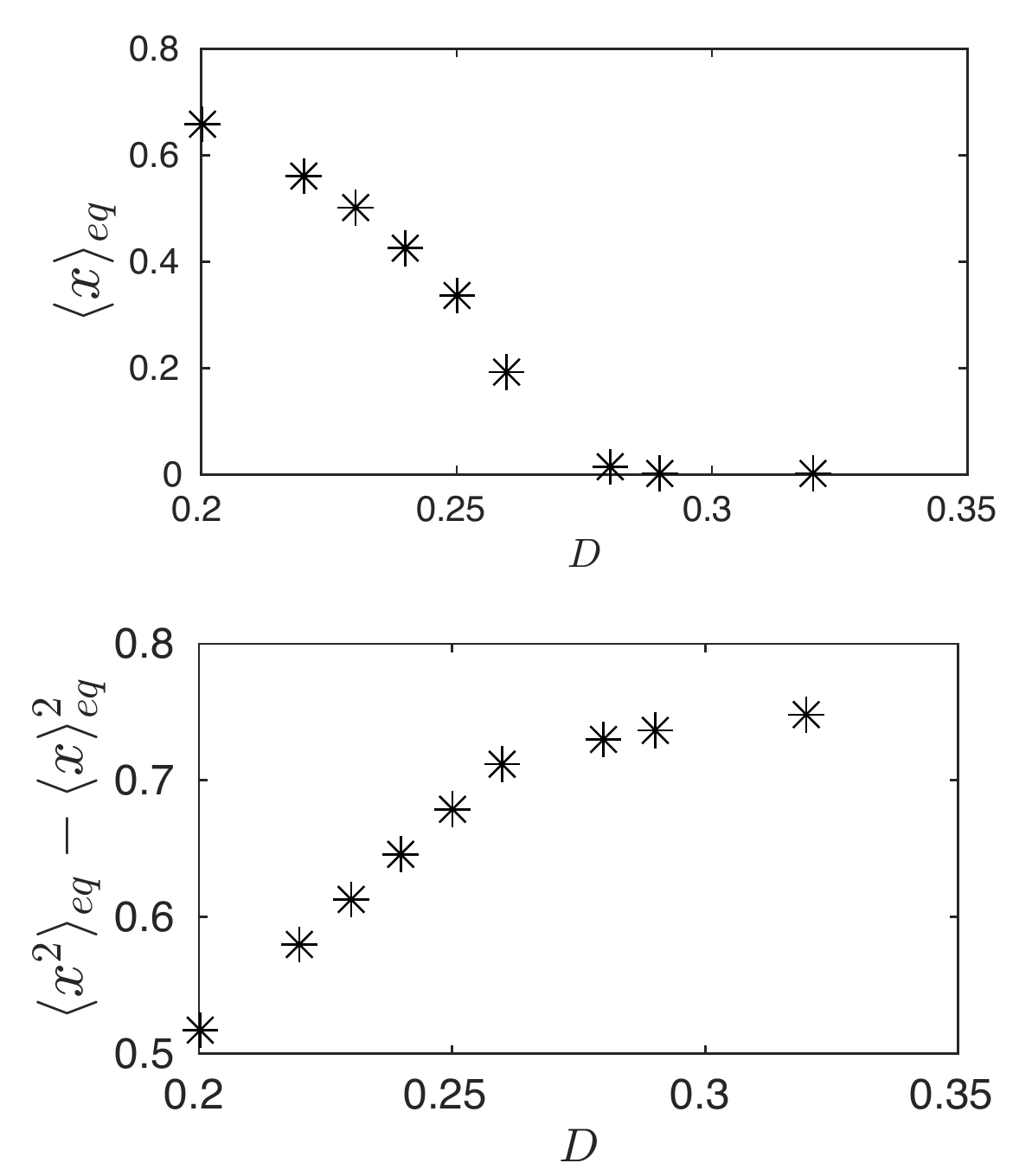,height=6cm}}
\caption{\label{xm}\small The first two equilibrium cumulants of the collective variable for 
$\theta=0.5$ and a range of noise values $D$, for a system with $N=11$.}
\end{figure}

In Figs. \ref{sm} and \ref{xm} we depict the behavior of the
equilibrium first and second cumulant moments of the collective variable and that of an individual variable 
 for a system with $N=11$ degrees of freedom and mean-field global dynamics. The parameter values are
  indicated in the captions. The average values
have been obtained from the corresponding numerical integration
involving the equilibrium histograms constructed from the Langevin
simulations. As expected from Eq.\ (\ref{cum}), the first moment of the individual and the collective variables are the same, 
while the second cumulant for the collective variable is smaller than that of the individual one. For the parameter values 
used in these figures, the probability distribution centered at $s=0$ (or $x=0$) is unstable for $D$ smaller than
a bifurcation value. In the figures we have just depicted that branch of the stable first moments
which are reached from the imposed initial condition $S(0)=0.3$.

\begin{figure}
\centerline{\epsfig{figure=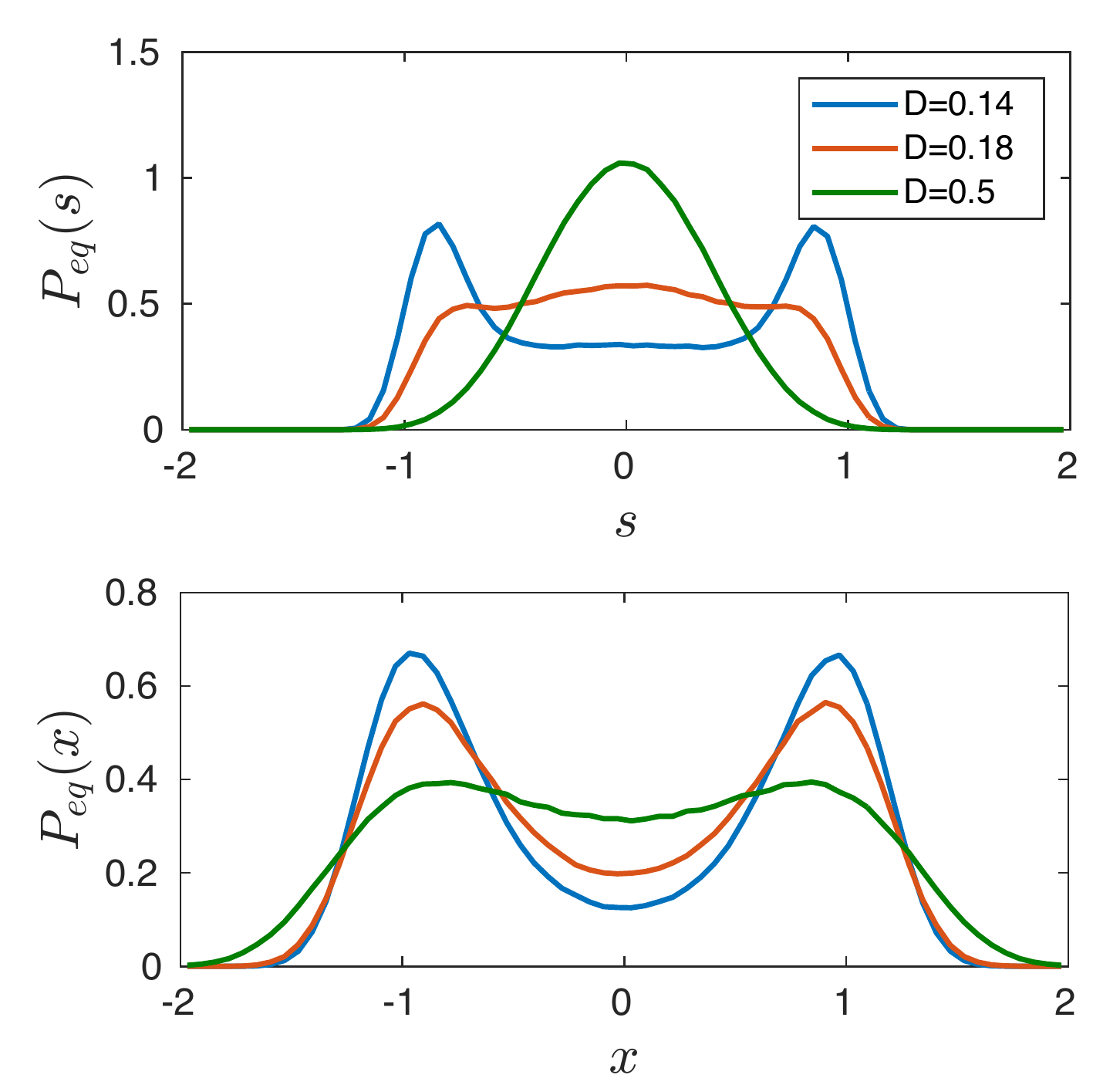,height=6cm}}
\caption{\label{NNFIN}\small $P_{eq}(s)$ (upper panel) and $P_{eq}(x)$ (lower panel) for a 
finite system with $N=11$ described by Eq.\ (\ref{NN1}), for 
$\theta=0.5$ and several noise values $D$}
\end{figure}

\begin{figure}
\centerline{\epsfig{figure=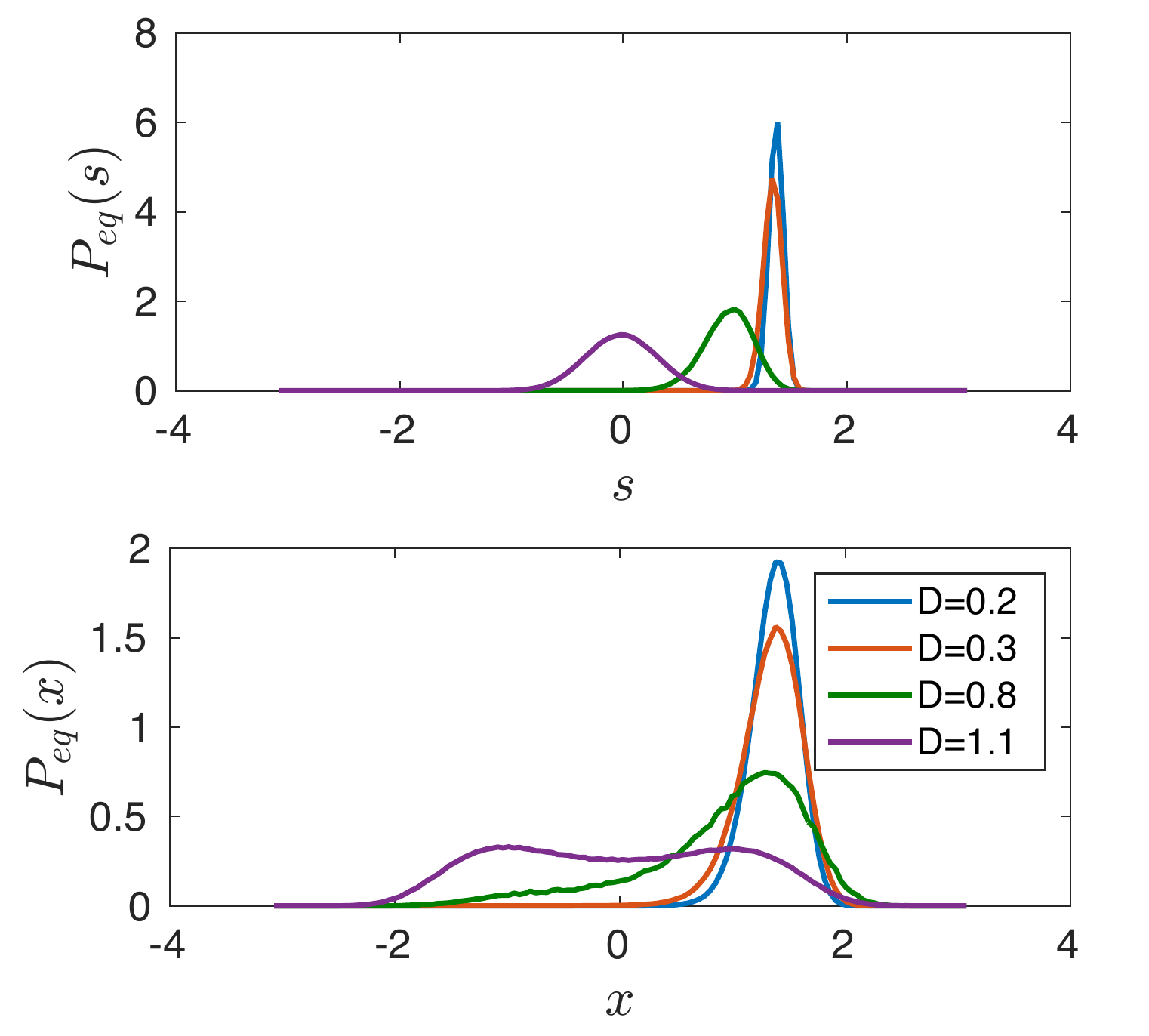,height=6cm}}
\caption{\label{NNMF}\small $P_{eq}(s)$ (upper panel) and $P_{eq}(x)$ (lower panel) for a 
finite system with $N=11$ and mean-field dynamics described by Eq.\ (\ref{EQ200}), for 
$\theta=0.5$ and several noise values $D$}
\end{figure}

The case of nearest neighbors interactions is qualitatively similar to the case of global interactions.
We will just depict as an example the contrast in the behavior of $P_{eq}(s)$ for the finite model in 
Eq.\ (\ref{NN1}) and the corresponding one with the mean field dynamics Eq.\ (\ref{EQ200}).
 In Fig. \ref{NNFIN} we plot the behavior of $P_{eq}(s)$ and $P_{eq}(x)$ for a system with $N=11$ and
  nearest neighbors coupling. In Fig. \ref{NNMF} the same distributions are sketched but now with the systems 
  described by the mean-field dynamics.
  In both dynamics 
  there is a change in the shape of the equilibrium distribution as the noise value is varied. But for 
  the finite case described by Eq.\ (\ref{NN1}) the equilibrium distribution is always unique regardless of the initial preparation. 
On the other hand, the mean field dynamics leads to a bifurcation of the distribution as $D$ is varied. 
For large values of $D$ there is a single stable $P_{eq}(s)$. As $D$ is decreased while keeping $\theta$ 
constant, the zero centered distribution becomes unstable and two stable distributions appear with peaks 
located at $s>0$ or $s<0$ depending on the initial preparation. 

\section{Conclusions}
We have carried out numerical simulations to study some aspects of the behavior of complex stochastic systems formed by a finite  number of subsystems with bistable intrinsic dynamics and mean field couplings. We have contrasted the behavior of each individual degree of freedom with that of a collective variable characterizing the entire system. 

The systems considered have finite sizes. Then, the collective variable is also a stochastic variable with a probability density with a finite width, in contrast with what occurs in the infinite size limit. Furthermore, the mean field dynamics is compatible with the possible existence of several probability distributions in some regions of the parameter space. There is a transition from a single stable distribution to the several stable ones as the noise strength is varied for a fixed value of the coupling parameter. In the case of multiple distributions, the one that is observed depends upon the initial preparation.

We argue that if the stochastic collective variable is the accessible one, and it is  such that its average value depends upon the initial conditions, the modelling of the subsystems dynamics requires the introduction of some sort of feedback  of the average collective behavior on the individual dynamics. An example of this feedback is the mean field dynamics considered in this paper. It should be pointed out that in order to have the above mentioned transition between one or several distributions, the intrinsic bistable dynamics of each individual degree of freedom is essential.

\end{document}